\documentclass{article}

\usepackage{amsmath,amssymb}
\usepackage{graphicx}
\usepackage{marvosym}

\usepackage[utf8x]{inputenc}

\usepackage{etoolbox}
\AtBeginEnvironment{quote}{\singlespacing\small}

\usepackage{textcomp,marvosym}

\newcommand{\beginsupplement}{%
        \setcounter{table}{0}
        \renewcommand{\thetable}{S\arabic{table}}%
        \setcounter{figure}{0}
        \renewcommand{\thefigure}{S\arabic{figure}}%
     }

\usepackage{cite}

\usepackage{nameref,hyperref}

\usepackage{microtype}
\DisableLigatures[f]{encoding = *, family = * }

\usepackage[table]{xcolor}
\usepackage{authblk}
\usepackage{array}

\begin{document}

\title{Towards a new social laboratory: An experimental study of search through community participation at Burning Man} 
\author[1*]{Ziv Epstein}
\author[2]{Micah Epstein}
\author[3]{Christian Almenar}
\author[1]{Matt Groh}
\author[1]{Niccolo Pescetelli}
\author[1,4]{Esteban Moro}
\author[1]{Nick Obradovich}
\author[1]{Manuel Cebrian}
\author[1]{Iyad Rahwan}
\affil[1]{MIT Media Lab}
\affil[2]{Rhode Island School of Design}
\affil[3]{Intrinsic.com}
\affil[4]{GISC \& Universidad Carlos III de Madrid, Spain}
\maketitle

%
%






\section*{Abstract}
The ``small world phenomenon,'' popularized by Stanley Milgram, suggests that individuals from across a social network are connected via a short path of mutual friends and can leverage their local social information to efficiently traverse that network. Existing social search experiments are plagued by high rates of attrition, which prohibit comprehensive study of social search. We investigate this by conducting a small world experiment at Burning Man, an event located in the Black Rock Desert of Nevada, USA, known its unique social systems and community participation. We design location-tracking vessels that we routed through Burning Man towards the goal of finding a particular person. Along the way, the vessels logged individual information and GPS data. Two of fifteen vessels made it to their designated people, but a month after Burning Man. Our results suggest possible improvements to limit rates of attrition through community participation and a design methodology that emphasizes cultural practices to aid social experimentation.

\section*{Introduction}
Popularized by Stanley Milgram in the 1960’s, the small world experiments were a series of investigations designed to measure how individuals in society are connected to one another (proxied by average path length) \cite{travers1967small}. The original experiment sought to understand how closely connected two randomly chosen individuals are to one another. In order to measure the closeness, the small world experiments tasked source individuals to send a letter to a target individual with the stipulation that they may only send the letter to people who they know on a first name basis, and who would in turn then forward the letter on towards the target. With a median path length of 5, these experiments suggested that society is a small world network - whereby any two individuals are connected by short paths. 

Social search has been recognized as an important aspect of modern society, as it allows groups of people to complete tasks they would not be able to do by themselves, such as finding red weather balloons in the infamous DARPA challenge \cite{tang2011reflecting, pickard2011time}. Since Milgram, there have been many similar small world experiments, such as Dodds et al, which replicated the finding using email chains instead of physical mail \cite{dodds2003experimental}. These experiments systematically suffered from high rates of attrition (no reported small world experiment has had an attrition rate lower than 70\%). Not only is social search an inherently difficult problem, but it seems that people are, on average, unwilling to engage in the cooperative behavior necessary to complete paths. 

This suggests that there are at least two critical ingredients for social search.  The first ingredient is the \textit{information we have about the network}, which manifests in efficient routing of information. The second ingredient is \textit{willingness to engage}, which manifests as low attrition. In most of the famous social search experiments discussed above, we had a large amount of information about the internet, email and social networks used, but there was a lack of willingness to participate. This lack of willingness to participate stems from the fact that these studies relied on people going out of their way to forward along the message (sending a letter in the mail and sending an email) without any tangible incentive.  

 The winning team of the Red Balloon Challenge used financial incentives to increase the willingness to participate, but still required people to go out of their way to engage \cite{tang2011reflecting, pickard2011time}. But how would social search fare in an an environment with a higher baseline willingness to participate, and where incentives to cooperate are hard-coded? 

\begin{table}
	\centering
\begin{tabular}{ |l|l|l|l| } 
 \hline
 Experiment &  \begin{tabular}{@{}c@{}}Knowledge of \\ the Network\end{tabular}&\begin{tabular}{@{}c@{}}Willingness to\\ participate\end{tabular} & \begin{tabular}{@{}c@{}}Network Size \\ (size during experiment)\end{tabular} \\  \hline \hline
 Milgram (snail mail) \cite{travers1967small} & +++++ & + & USA(300M) \\ \hline
 Dodds et al (email) \cite{dodds2003experimental} & ++++ & ++ & World (7B) \\ \hline
 Red Balloon \cite{pickard2011time, tang2011reflecting} & +++ & +++ & USA (300M) \\ \hline
 Burning man & + & +++ & Black Rock City (70K) \\ \hline
\end{tabular}
 \caption{Hypothesized dynamics of social search.}
 \label{tab:equation} 
\end{table}
One such environment is Burning Man, an annual event that takes place in the temporarily constructed city of Black Rock City, north of Reno, Nevada. It is an art and community centered experiment grounded in  principles such as decommodificaiton, leave no trace and communal effort \cite{kozinets2002can}. Devoid of technology, money and the standard rhythm of urban life, can the event serve as an \textit{in situ} laboratory to study how human sociality functions in harsh, low-information, environments? In many ways, this unique environment mirrors many of the attributes of historical human civilizations, where many social norms and intuitions such as cooperation evolved \cite{bowles2002prosocial,chudek2011culture, bowles2003origins}. 

Burning Man also has a unique cultural context. A core aspect of the event is the notion of participatory engagement. The 6th principal of Burning Man, Communal effort, states: ``Our community values creative cooperation and collaboration. We strive to produce, promote and protect social networks, public spaces, works of art, and methods of communication that support such interaction.'' This unique ethos speaks to Milgram's original vision of a connected, collaborative world. 

The unique dynamics and culture of Burning Man have been studied by anthropologists, sociologists and other scholars for decades.\footnote{See the Burning Academics page (\url{https://burningman.org/culture/philosophical-center/academics/}) or the Burning Nerds group for an exhaustive list} This work has ranged from ethnographic studies of Burning Man practices \cite{hockett2005participant, rosenbloom2017aural} to organizational theory to as how the large-scale participatory network emerged and has been sustained \cite{chen2012laboring, chen2009enabling, hoover2008realizing}. As far as quantitative social science, Crocket, Yudkin and colleagues, in collaboration with the Burning Man Census, have empirically investigated the transformational experiences people undergo at Burning Man \cite{trans1, trans2}. The Burning Man Census team has also measured various sociodemographic, ethnoracial and social data. \footnote{For more information about Burning Man demographics, see the work of the Black Rock City Census \url{https://drive.google.com/file/d/1hbZtR38TiEqDgA28STIFwS5Ae1r4WrYP/view}} Despite this rich literature, there is still many open questions about how the mechanisms of Burning Man foster scalable cooperation.

To investigate this, we conducted a small world experiment at the 2018 Burning Man, in which we routed a set of 15 “vessels” across the event. These vessels each contained information about a particular individual at Burning Man this year. These are individuals who volunteered en masse to help us with the project (over 350 people volunteered to participate in the project). The vessels, through cooperative handoffs and community networks,  moved across Black Rock City. Along the way, these vessels were annotated with information about their journey, so we could retrace and understand the patterns of their voyage.

\section*{Materials and methods}
\subsection*{Terminus Selection}
We recruited individuals attending the 2018 Burning Man event to be ``terminii'' of the social search (analogous to Milgram's ``target people''). Following a recruitment post in the Burning Man Journal, 361 volunteers visited our website and each wrote short descriptions about themselves, and provided general demographic information.  We selected 15 Terminii from this pool using the following criterion. First, we performed stratified random sampling to select a random individual  \Gentsroom. In the hopes of approximating a representative sample, we stratified individuals on nationality and the means by which they found out about the project in order to better approximate the international and heterogeneous nature of the Burning Man social network. We then selected \Gentsroom \text{ }as a Terminus if \Gentsroom\text{ } has not already been selected, and their description provides enough information such that social search is possible, but not too much information so that finding \Gentsroom\text{ } is trivial.

In particular, since each handoff should bring the vessel ``closer'' to the Terminus, if \Gentsroom's description contained no useful information by which the vessel can get closer to that person, social search would fail. Alternatively, if \Gentsroom's description contains too much information (i.e. exactly where they camp), then social search would be trivial. Table~2 shows descriptions from some of the selected Terminii. We repeat this sampling process until all 15 Terminii have been selected.

\begin{table}
	\centering
\begin{tabular}{ |c|p{9cm}| } 
 \hline
 Terminus ID & Description \\  \hline

 1 & I am an entomologist who looks for bugs with my net on the playa.  I often have wings and antennae.\\ \hline
 4 & I'm typically up all night, early morning and late afternoon, spending the majority of my nights in \\
  & deep playa at Mayan Warrior. I'm always wearing nipple tassels, septum piercings and bright colours- \\
  & usually one of the colours from the rainbow from head to toe.\\
 \hline
  8 &  I volunteer as a Ranger, often watch the sunrise from deep plays and may have a flame thrower.  I do  have\\
   &  a weakness for iced coffee and a pithe helmet with a foxtail and camps at a place that washes hair - its \\
   & a cookie monster art car  \\  \hline
\end{tabular}
\label{tab:desc}
\caption{Description of Selected Terminii}
\end{table}
\subsection*{Vessel Components}
\begin{figure}
\includegraphics[width=\textwidth]{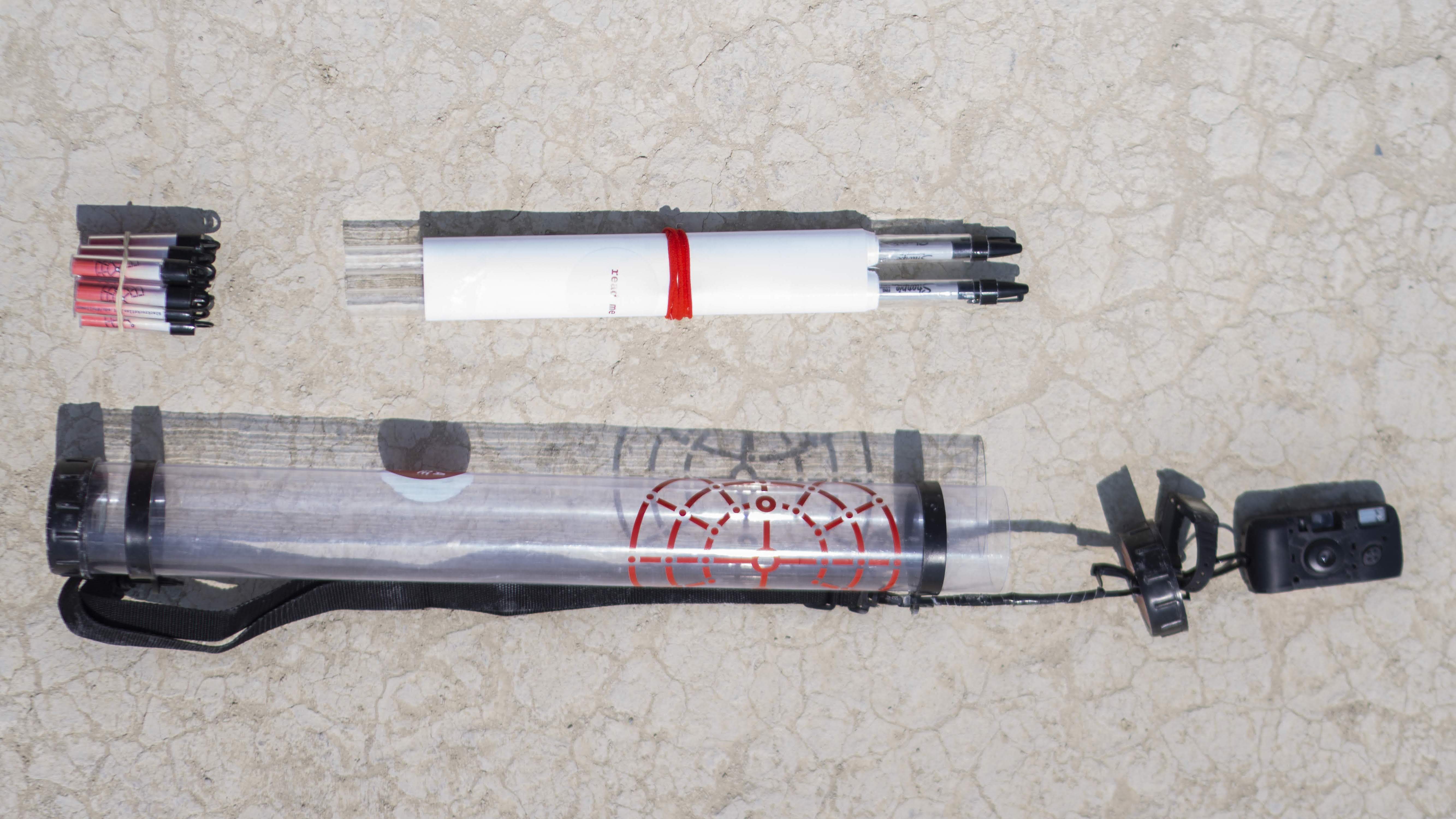}
\caption{{\bf Contents of vessel.} Each vessel contained a GPS tracker, a scroll, gifts, a disposable camera, two pens, and a watch}
\label{fig1}
\end{figure}

\begin{figure}[!h]
\includegraphics{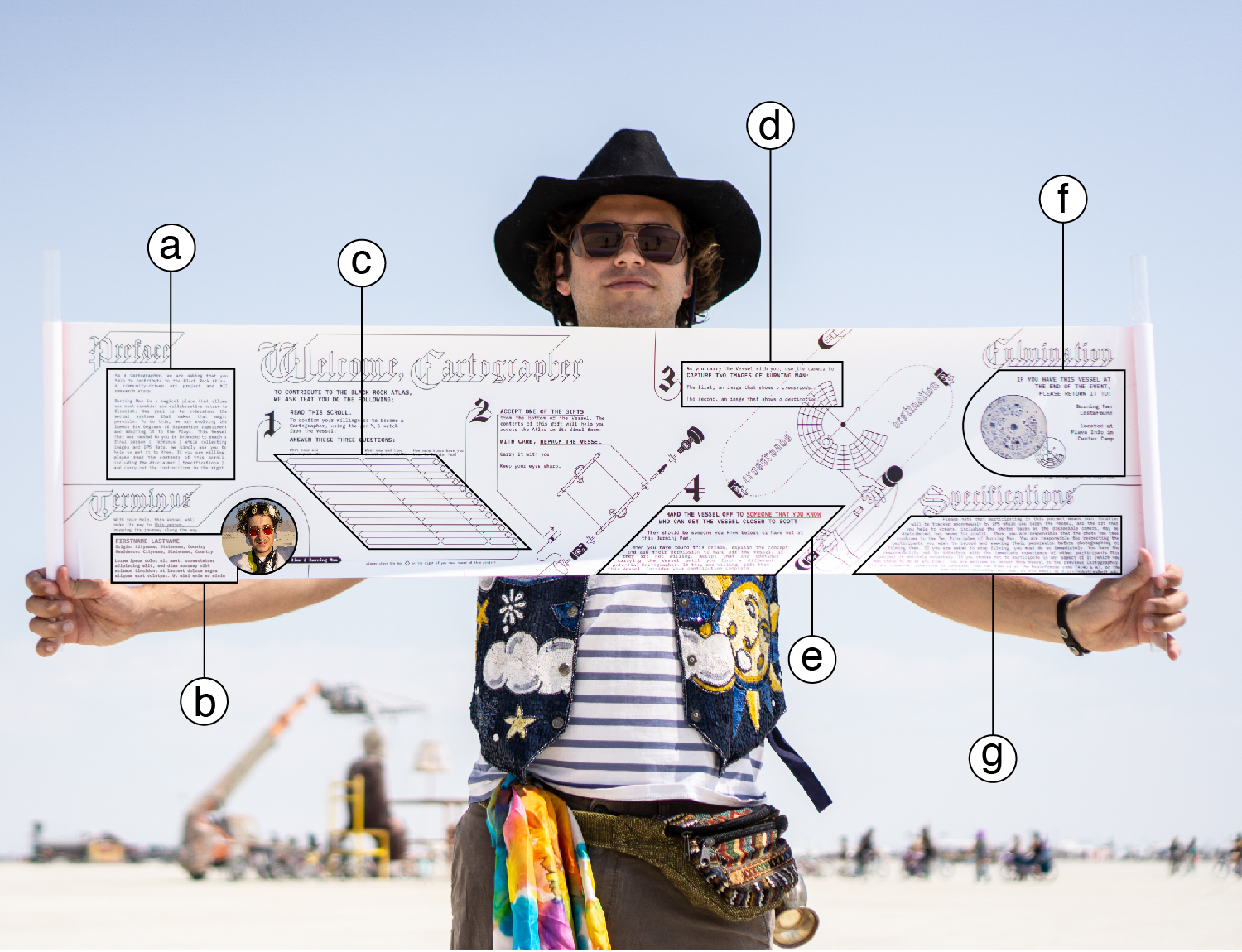}
\caption{{\bf The scroll.}
Contains informed consent context (a), Terminus information (b), table for informed consent affirmation and hand-off time entry (c), instructions for photography (d), specific hand-off procedure (e), drop-off location if the event ends (f), and necessary disclosures and contact information (g).}
\label{scroll}
\end{figure}

In designing the vessels, we made sure to balance the functional requirements of the experiment with a design methodology that emphasized \textit{emic} Burner traditions and practices \cite{headland1990emics}. This culturally specific design approach came together as a cohesive and unconventional visual presence that we tailored to invite enthusiasm and participation from our participants. Some key examples of this are as follows: 

\textbf{Scroll:} The key information that we identified for this study included necessary context and disclosures, terminus information, and instructions for participation. We chose to present this information in a scroll to align with the spiritual themes that Burning Man is associated with\cite{gilmore_2010}. Further, that key information was presented in a way that would align it less with its “default” \footnote{“Default: The rest of the world that is not Black Rock City during the Burning Man event”\cite{burningMan}} context and associate it more with the rituals that are common place at burning man\cite{gilmore_2010}. For instance, “participant” was replaced with “Cartographer.” See the top right of Figure~\ref{fig1} or Figure~\ref{scroll}  for an image of the scroll we used.

\textbf{Gifts:} The exchange of gifts is a central tenant of Burning Man culture \cite{kozinets2002can}. Indeed, Burning Man is a ``gift economy,'' whereby valuables are neither sold or traded, but are rather given with no explicit agreement for reciprocity. We decided to leverage this characteristic to make the user experience engaging (and thus increase the probability of handoff). A consistent manifestation of this are the ``playa pendants''\cite{raiser_2017} that decorate the necks of most attendees to Burning Man. We deferred to this culture by ``gifting'' our own interpretation of a playa pendant to each participant. In addition to aligning our study with that facet of Burning Man culture, the gifts served a two-fold functional purpose: it contained the link to a follow-up survey to collect valuable feedback while spreading the visual presence of the study across the event. See the top left of Figure~\ref{fig1} for an image of the gifts.

\textbf{Disposable Camera:} the camera was included to allow for the study to have a participatory and creative facet, which is a well-documented method of generating social capital and involvement\cite{dekker2003social}. See the bottom right of Figure~\ref{fig1} for an image of the camera.

All components above were assembled in a poster tube that we identified to be durable and portable enough to passed around the event for a week, as well as a SpyTec GL3000 GPS Tracker, and pens and a watch for participants to affirm their consent and record their responses and the hand-off time. Various components were attached, reinforced, or relabelled to improve durability or clarity. 

\subsection*{Experimental Procedures}
On August 27, 2018, we brought 15 vessels to Center Camp and Esplanade in Black Rock City for the hand-offs. We explained the study to curious passersby, including the idea of six degrees of separation, the nature of the experiment, the responsibilities associated with participation,and finally asked if they were interested in participating. If so, we had them sign the scroll which served as a consent form, gifted them the first pendant from the vessel, and asked that they explain the same details to whoever the hand the vessel off to\footnote{All of this information was also detailed in the scroll in case a participant forgot any details.}. Finally, we sent them on their way with vessel in hand.
\begin{figure}[!h]
\includegraphics[width=\textwidth]{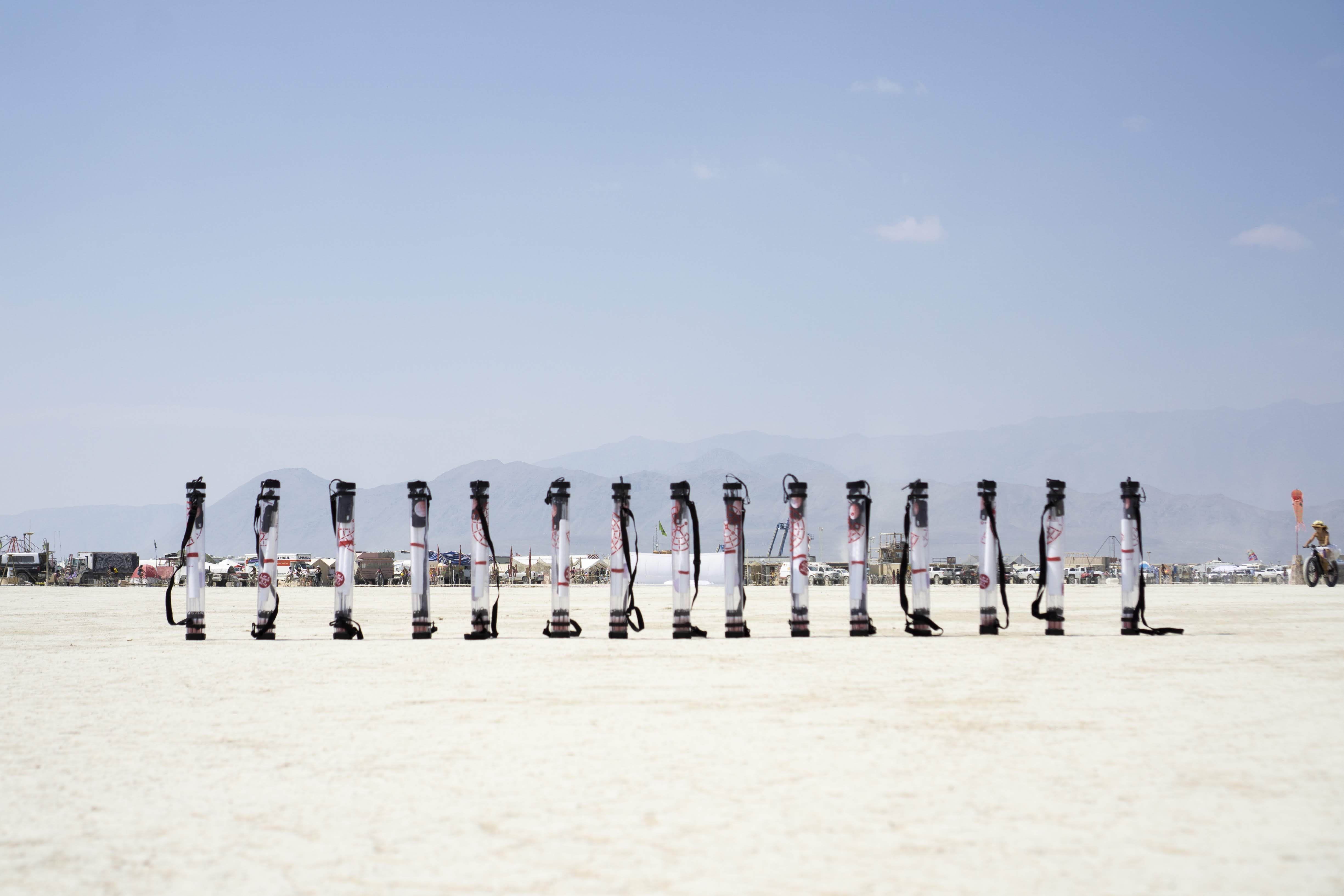}
\caption{{\bf Vessels at hand-off.} The vessels were arranged in a line to incite curiosity in passerby.}
\label{fig3}
\end{figure}
\subsection*{Vessel Collection}
The scrolls stipulated that the vessels should be returned to the Burning Man Lost and Found if either 1) the vessel found itself to the terminus, or 2) the event was over. We intermittently checked in with the Lost and Found to observe the state of the experiment, and collected vessels from the Lost \& Found as they were found there. We ascertained the results of the experiment by looking at the collected scrolls and following up with each of the 15 Terminii to see if they got their vessel. We were unable to analyze data from uncollected scrolls. For the scrolls that failed to be collected, we included contact information on our website and in the scroll itself that allowed individuals to get in touch with us if they found them after the event. 
\section*{Results}
\subsection*{Theoretical and experimental priors on search success}
Prior to the actual social search experiment, we first report two estimates on the probability of success for the experiment. 

The first estimate comes from a simulation framework with simple assumptions about the social search transmission. We assume the social network of Burning Man is a scale-free network with 70,000 nodes, parameterized by $\alpha, \beta, \gamma$.

\begin{figure}[h]
\includegraphics[width=0.49\textwidth]{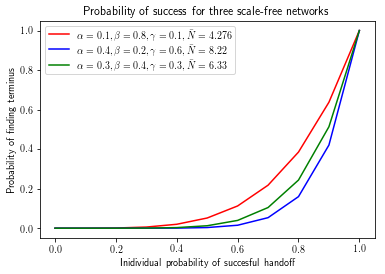}
\includegraphics[width=0.49\textwidth]{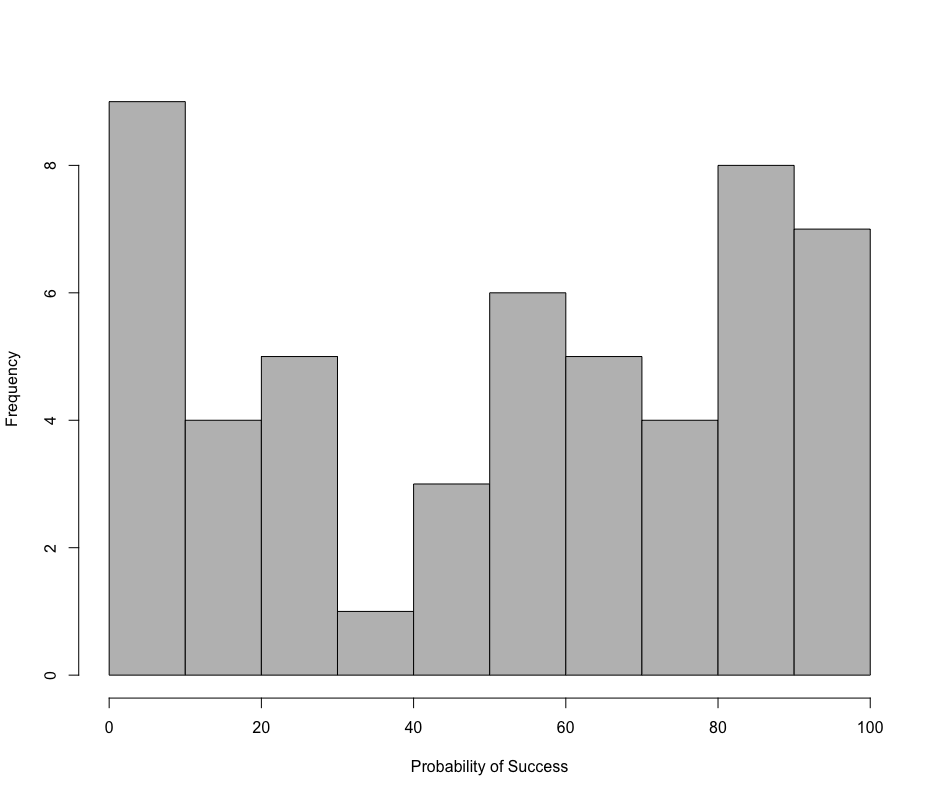}
\caption{{\bf Estimates of probability of success} Left: Theoretical model of success probability for three scale-free networks, depending on the individual probability of a successful handoff.  Right: distribution of reported probability of success of 55 Burning Man participants. }
\label{fig2}
\end{figure}

For a given $\alpha, \beta, \gamma$, we randomly select 500 pairs of nodes and compute the shortest path between them.  We do a grid search over $\alpha,\beta,\gamma \in [0.1,0.2,\cdots, 0.8,0.9]$ constrained by $\alpha+\beta+\gamma =1$ and we select three sets of parameters that correspond to the miminal, median and maximal average shortest path length, which are  \textcolor{red}{$\alpha_1 = 0.1, \beta_1 = 0.8, \gamma_1 = 0.1$}, \textcolor{blue}{$\alpha_2 = 0.4, \beta_2 = 0.2, \gamma_2= 0.6$}, and \textcolor{green}{$\alpha_3=0.3, \beta_3=0.4, \gamma_3 = 0.3$} respectively. We model the probability of success $p_i = k^{\bar{N_i}}$, where $k$ is the probability that a given person hands the vessel to the next, and $\bar{N_i}$ is the average shortest path length of graph $i$. Figure~\ref{fig2} (left) shows this $p_i$ over the unit interval of possible values for $k$. As the figure indicates, the individual probability of a successful hand-off needs to be quite high before social search becomes possible. 

The second estimate comes from a survey of 361 Burning Man participants, asking them about their perception of the likelihood of success. On being asked on the probability of success, 55 participants replied with a mean probability of 55.9\%. This distribution of reported percentages is shown in Figure~\ref{fig2}, does not look meaningfully different from a uniform distribution. Together, these two priors suggest that finding the terminus may be very difficult.

Table 2 shows the results from each vessel (all blank entries correspond to information we do not know). 

\begin{center}

\begin{tabular}{ |c|c|c|c|c| } 
 \hline
 Vessel ID & Location & \begin{tabular}{@{}c@{}}Successfully reached\\ terminus\end{tabular}& GPS Data & \# of handoffs \\  \hline
 1 &Unknown &   & & \\
 2 &Unknown &  & & \\
 3 &Reclaimed &  & Yes& 1 \\
 4 &Unknown && & \\
5 & Reclaimed && No & 4 \\
6 &Unknown && & \\
7 &Unknown && & \\
8 &Reclaimed &  & Yes &4 \\
9 &Unknown & & & \\
10 &Unknown & & & \\
11 & Reclaimed & \checkmark  & No&6 \\
12 &Reclaimed & \checkmark   & No&4 \\
13 &Unknown && & \\
14 &Unknown & & & \\
15 & Reclaimed && No & 1 \\
 \hline
\end{tabular}

\end{center}
We reclaimed three vessels from the Lost and Found over the course of the event, and two more from the Lost and Found immediately after clean up. After assessing the results from these vessels (\#3, \#5, \#8, \#11 and \#15), it seemed that no termini received their vessel. Indeed three vessels were found at the Lost and Found, but were turned in by individuals other than the termini. Of these three, two demonstrated that hand-offs did occur, while the third did not.

However, over a month after Burning Man, the Decompression events of Burning Man facilitated two vessels (\#11 and \#12) reaching their terminii. Decompression is a set of regional events held one to two months after Burning Man which enables the community to reconvene and cope with the re-entry to the default world \cite{gilmore2005afterburn}. For vessel \#12, we were contacted by a terminus who was given his vessel at the Denver Decompression Even on October 12, 2018. A fellow Denver Burning Man participant reached out to him on Facebook prior to the event, and they met up at the event for the final handoff. 

We had picked up the unfound vessel \#11 from the Lost and Found after Burning Man, and brought it to our local New England Decompression event, which  occurred in Providence, Rhode Island on November 3rd. Serendipitously, Terminus \#11 was at this event and our paths crossed. We gave them their Vessel and completed the chain as penultimate nodes. We did not know they would be there, and had no intention of completing chains, but  -- as is a canonical phrase at Burning Man -- the Playa provides.

 These two success handoffs were not contained to the temporal bounds of Burning Man, as we had originally hypothesized, and required further experimenter and technological intervention (for vessels \#11 and \#12, respectively), to succeed. However, in both cases, not only was social search successful, but the vessels were shepherded across state lines in a multi-week voyage.
\begin{figure}[h]
\includegraphics[width=0.99\textwidth]{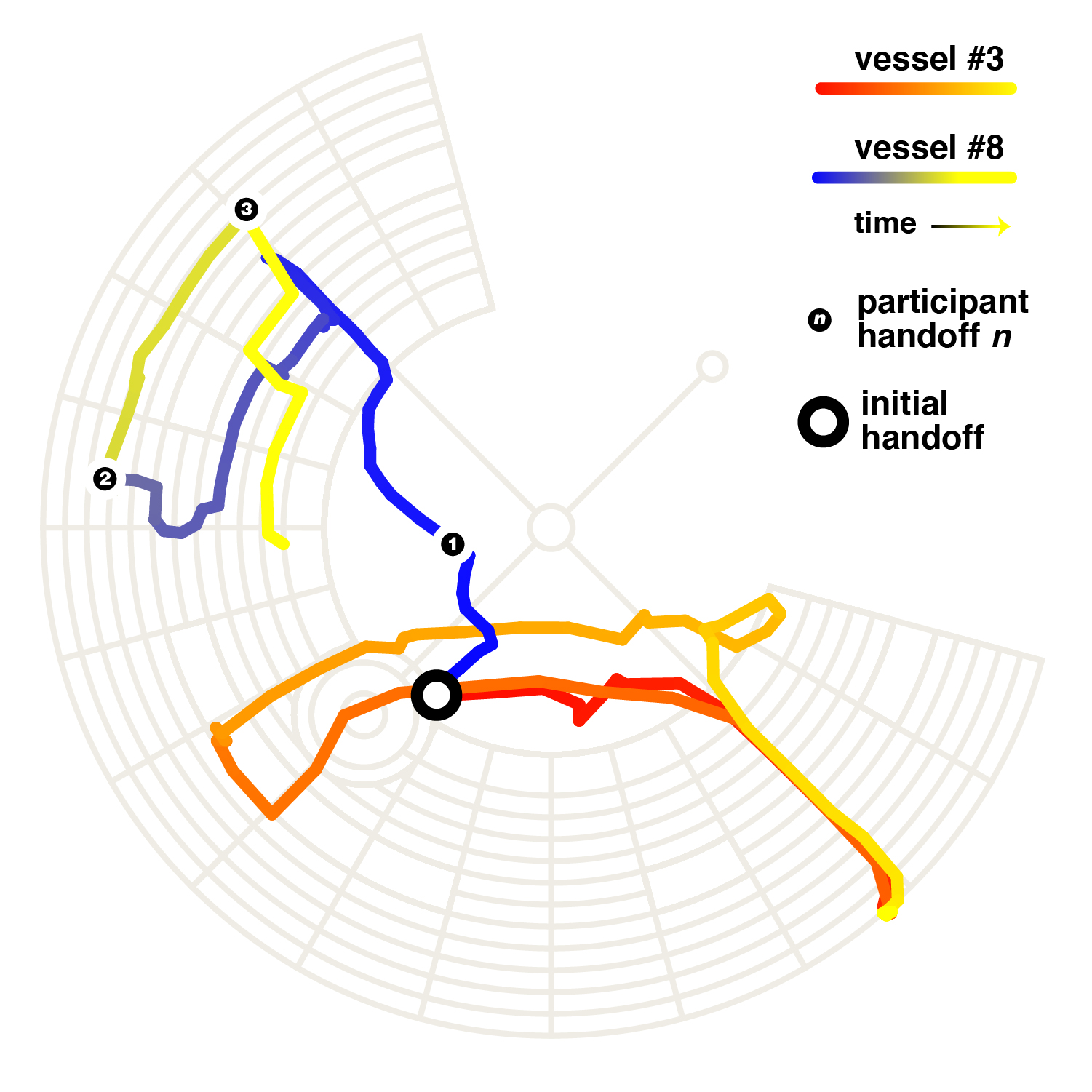}
\caption{{\bf Temporospatial data from vessels 3 and 8}}
\label{4}
\end{figure}
\subsection*{GPS Data}
The GPS system we used requires cell towers to broadcast a signal. Thus, since Burning Man had over 70,000 people, they were not able to dynamically broadcast their location, and ran out of battery after a week. The GPS units we were able to recover (\#3, \#8, \#11) were able to be charged and turned back on, which allowed their signal to broadcast. Vessel \#11 stored only a single GPS data point, which prevents us from extracting data from it.  Figure~\ref{4} shows the GPS data over vessels 3 and 8 over time. The remaining GPS devices failed to record data. 

The mobility patterns for vessels 3 and 8 show movement across both the open Playa (the center circle) and Black Rock City (the surrounding grid system). 

\section*{Discussion}
For two vessels social search was successful but in a way that we did not anticipate. In addition to these vessels success there is evidence that other vessels got close. One Terminus, a Black Rock Ranger, said that his vessel found its way to the Ranger headquarters and that he received a message to pick it up. When he went to retrieve it, however, the rangers could not find the item. When they eventually found it Monday afternoon, he had already left Black Rock City. These anecdotes suggests that future social search may be possible, and that more than 15 vessels may be necessary to measure social search.

The successful handoffs at the Decompression events suggest that the Burning Man social network is not as temporally and spatially bounded as we first thought. Rather, it is a globally distributed community with a long time horizon.

Furthermore, the overwhelmingly positive response to the project from participants and volunteers suggest that our design approach successfully generated active participation and community support. The 361 volunteers exceed our expectations, and yielded affirming responses such as ``Super excited for your project! I did my PhD on network community structure. Would be thrilled to help assemble data and analyze. I camp with the science based camp the phage. If you haven't been contacted about doing a talk about your project at one of our nightly sessions, please let me know!” and “Connections are the threads of life, so I look forward to the ones that bring people to me and bring life to this project!'' 
In addition, one of our volunteers was a coordinator of the official Burning Man Lost \& Found. This allowed us to use their facilities as a repository for vessels at the end of the event, and also reinforced our assumptions about the existence and outcomes of active participation at Burning Man. 

Our study, both as an exploration of social search and participatory engagement, has a couple limitations. First, our selection process of individuals biases our results towards success. Second, our small sample size of only 15 vessels limits our ability to reason about an effect, or account for variance. Finally, the Burning Man population deviates from a representative population which makes assessing the external validity of experiments conducted at the event difficult and an open future research question \cite{census}. In addition, because of the costly and specific nature of the experiment, reproducibility is inherently difficult. 

Future work includes scaling up the number of vessels to better quantify an effect and using a more comprehensive and representative recruiting scheme. There are also many unexplored questions about how to design streamlined and compelling user experiences to increase the probability of participant hand off. In addition, one could imagine improving the robustness of the GPS system with local data storage and augmented network access towards conducting large human mobility studies in extreme environments. 

We hope this work encourages others to explore new design innovations in the social search problem space. In addition, we hope this work inspires social and information scientists to consider Burning Man a new laboratory, where other interesting hypotheses and social dynamics can be studied and mapped. 

\beginsupplement

\section*{Acknowledgments}
The authors would like to thank Stewart Mangrum and Caveat Magister at the Burning Man Organization for providing institutional resources. They would also like to thank Sarah Newman, Kim Albrect, bunnie Huang and Jie Qi for invaluable discussion and feedback. They would also like to thank the fifteen anonymous terminii who participated in the project, and the Burning Man community for their continued support and enthusiasm. 


\section*{Author Contributions}
ZE, CA and MC conceived the original idea. ZE and ME conducted fieldwork and analyzed the data. CA conducted fieldwork and helped interface with the Burning Man community. MG and NP helped with technical and experimental details. EM and NO analyzed data, contributed to the research design and helped with technical and experimental details. MC helped interface with the Burning Man community. IR contributed to the research design.

%
%
%
\bibliographystyle{abbrv}
\bibliography{bmbib}
\end{document}